\documentclass[pra,aps,twocolumn,superscriptaddress,showkeys,nofootinbib]{revtex4-1}
\usepackage[latin1]{inputenc}
\usepackage{amsmath,amsthm,amssymb,graphicx,subfigure,xcolor}
\usepackage{tikz}
\usepackage{mathrsfs}
\usepackage{graphics,float}
\usepackage{epstopdf}
\usepackage{color}
\usepackage[unicode=true]{hyperref}
\usepackage{booktabs}
\usepackage{tabularx}
\usepackage{tabu}
\usepackage{multirow}
\allowdisplaybreaks[4]
\hypersetup{
     colorlinks=true,       		% false: boxed links; true: colored links
     linkcolor=blue,          	% color of internal links
     citecolor=blue,            % color of links to bibliography
     urlcolor=blue,           	% color of external links
 }
\newtheorem*{theorem*}{Theorem}

\newtheorem*{corollary*}{Corollary}

\newtheorem*{lemma*}{Lemma}

\newtheorem*{proposition*}{Proposition}
\theoremstyle{definition}

\newtheorem*{definition*}{Definition}
\theoremstyle{remark}

\newtheorem*{remark*}{Remark}
\begin{document}
\title{Quantifying multipartite quantum states by ($k+1$)-partite entanglement measures}
\author{Hui Li}
\author{Ting Gao}
\email{gaoting@hebtu.edu.cn}
\affiliation{School of Mathematical Sciences, Hebei Normal University, Shijiazhuang 050024, China}
\author{Fengli Yan}
\email{flyan@hebtu.edu.cn}
\affiliation{College of Physics, Hebei Key Laboratory of Photophysics Research and Application, Hebei Normal University, Shijiazhuang 050024, China}
\begin{abstract}
In this paper, we investigate how to quantify the quantum states of $n$-particles from the point of $(k+1)$-partite entanglement $(1\leq k\leq n-1)$, which plays an instrumental role in quantum nonlocality and quantum metrology. We put forward two families of entanglement measures termed $q$-$(k+1)$-PE concurrence $(q>1)$ and $\alpha$-$(k+1)$-PE concurrence $(0\leq\alpha<1)$, respectively. As far as the pure state is concerned, they are defined based on the minimum in entanglement. Meanwhile, rigorous proofs showing that both types of quantifications fulfill all the requirements of an entanglement measure are provided. In addition, we also propose two alternative kinds of entanglement measures, named $q$-$(k+1)$-GPE concurrence $(q>1)$ and $\alpha$-$(k+1)$-GPE concurrence $(0\leq\alpha<1)$, respectively, where the quantifications of any pure state are given by taking  the geometric mean of entanglement under all partitions satisfying preconditions. Besides, the lower bounds of these measures are presented by means of the entanglement of permutationally invariant (PI) part of quantum states and the connections of these measures are offered. Moreover, we compare these measures and explain the similarities and differences among them. Furthermore, for computational convenience, we consider enhanced versions of the above quantifications that can be utilized to distinguish whether a multipartite state is genuinely strong $k$-producible.
\end{abstract}

%\pacs{03.65.-w, 03.65.Ta}

%\keywords{Suggested keywords}

\maketitle
%\tableofcontents{}

\section{Introduction}
Quantum entanglement, being a crucial physical resource, manifests obvious superiorities ranging from quantum cryptography \cite{7,8,9,10}, quantum teleportation \cite{11,12,13}, to quantum communication \cite{14,15,6}, as compared to the classical theory. One of the key issues in entanglement resource theory is the development of measures to quantify the entanglement of quantum states, which has significant implications in both theoretical studies and practical applications.

Bipartite quantum states, as the simplest case, are either entangled or separable. And the study of entanglement measures for bipartite systems has been followed by successive achievements, such as concurrence \cite{22,23,24,25}, negativity \cite{26,27}, logarithmic negativity \cite{51}, entanglement of formation \cite{28,29}, parametrized concurrence \cite{5,16}, which can tell whether a quantum state is entangled.

A quantum state of $n$-particles can be entangled in a variety of ways, however, leading to an exponential growth in the complexity of entanglement structures of $n$-partite quantum states as the number of particles increases. It was realized that multipartite quantum states can be classified from different aspects. Among them, several relevant measures \cite{20,19,40,41,42} and detection criteria \cite{43,44,45,46,47,48} on how to unambiguously determine whether a quantum state is genuinely multipartite entangled were introduced. In addition, an $n$-partite quantum state may be partially entangled rather than genuinely multipartite entangled. For the sake of understanding the whole hierarchy of multipartite states more precisely, one discussed them in the light of two different directions. On the one hand, much effort has been devoted to characterizing $n$-partite quantum states $\rho$ in terms of $k$-separability and $k$-nonseparability $(2\leq k\leq n)$ \cite{3,4,18,49,50,17,21}, reflecting the question of how many partitions are separable. On the other hand, the quantum states of $n$-particles can also be divided from the point of $k$-producibility and $(k+1)$-partite entanglement $(1\leq k\leq n-1)$ \cite{1,2,17,30}, which answers the question of how many particles are entangled.

The $k$-separability and $k$-producibility stemmed from multipartite quantum states are deemed to be instrumental in quantum information theory, including in spin chains \cite{1,2} and in quantum nonlocality \cite{31,32,33,34}. In addition, $k$-producibility can be applied to quantum metrology \cite{35}. For larger $k$, the $k$-producible entanglement states have higher sensitivity and hence higher accuracy in phase estimation, which has been demonstrated experimentally \cite{36,37,38} and has also resulted in the emergence of $k$-producible entanglement criteria \cite{39}. It is acknowledged that entangled quantum states are ubiquitous as a resource, and their utility in these applications usually relies on how entangled the quantum states are. Therefore, our aim is to characterize and quantify the entanglement of quantum states from the perspective of $(k+1)$-partite entanglement.

The paper is organized as follows. In Sec. \ref{II}, we cover some of the necessary basics. In Sec. \ref{III}, we present two kinds of entanglement measures, called $q$-$(k+1)$-PE concurrence $(q>1,~1\leq k\leq n-1)$ and $\alpha$-$(k+1)$-PE concurrence $(0\leq \alpha< 1,~1\leq k\leq n-1)$, respectively, where the quantifications of any pure state are presented by adopting the method that takes the minimum in entanglement. Simultaneously, we show rigorously that these quantifications obey all conditions to be an entanglement measure, including entanglement monotone, strong monotone, vanishing iff the quantum state is $k$-producible, convexity, subadditivity. In Sec. \ref{IV}, two other families of alternative  entanglement measures, termed $q$-$(k+1)$-GPE concurrence and $\alpha$-$(k+1)$-GPE concurrence, are put forth in terms of the geometric mean of entanglement. Moreover, the lower bounds on these measures are derived and the links between them are established. In addition, these measures obtained by us are compared and the similarities and differences among them are revealed in Sec. \ref{V}. Furthermore, the modified versions of the above quantifications for arbitrary quantum states are rendered in Sec. \ref{VI}, which can be harnessed to determine whether a quantum state is genuinely strong $k$-producible and are computationally convenient. We summarize the main conclusions in Sec. \ref{VII}.

\section{Preliminaries}\label{II}
We embark on introducing some basics that can be used in the subsequent sections, and state that the ranges of $q$, $\alpha$, and $k$ are, respectively, default to $q>1$, $0\leq\alpha<1$, and $1\leq k\leq n-1$ in the absence of annotations.

An $n$-partite pure state $|\phi\rangle$ is called $k$-producible ($1\leq k\leq n-1$) \cite{1,2,17} if there exists some splitting $A=A_1|A_2|\cdots|A_m$ (of set $S=\{1,2,\ldots,n\}$) such that $|\phi\rangle=\otimes_{t=1}^{m}|\phi_t\rangle_{A_t}$ and $|A_{t}|\leq k$, where $|A_{t}|$ denotes the number of particles of subsystem $A_t$, the partition must satisfy simultaneously the conditions: (a) the union of $A_1,A_2,\ldots,A_m$ amounts to the set $S$, (b) any pair is disjoint. An $n$-partite mixed state $\rho$ is referred to as $k$-producible if it can be expressed as $\rho=\sum_ip_i|\phi_i\rangle\langle\phi_i|$, where $|\phi_i\rangle$ could be $k$-producible under different partitions. Otherwise, the quantum state is termed $(k+1)$-partite entangled. In fact, $k$-producible reflects the nature of ``how many particles are entangled".

A pure state $|\phi\rangle$ is said to be genuinely $k$-producible (or genuinely $k$-partite entangled) \cite{2} if it is not $(k-1)$-producible, meaning that for $|\phi\rangle=\otimes_{t=1}^{m}|\phi_t\rangle_{A_t}$ and $|A_{t}|\leq k$ there is at least one state $|\phi_t\rangle_{A_t}$ whose the quantity of particles is equal to $k$ and which cannot be factorizable. One thinks that a mixed state $\rho=\sum_ip_i|\phi_i\rangle\langle\phi_i|$ is genuinely $k$-producible if there is at least one $|\phi_i\rangle$ is genuinely $k$-producible.

Note that $m$ is varied under different partitions. For example, we determine whether a 5-partite quantum state is 2-producible, the number of each subsystem to be considered might be $2|2|1, 2|1|1|1, 1|1|1|1|1$ corresponding to $m=3,4,5$, respectively. And there is a relation between $n$, $m$, and $k$, that is, $m\geq n/k$.

Remarkably, the set of $n$-separable quantum states is identical to the set constructed by quantum states of $1$-producible. A quantum state is called genuinely multipartite entangled if it is not $(n-1)$-producible. The set of $(n-1)$-separable quantum states, however, is a subset of the set of $2$-producible quantum states. Let $P_k$ $(k=1,2,\ldots,n-1)$ denote the set of $k$-producible quantum states and $P_n$ be the set of all quantum states, the relation among $P_1,P_2,\ldots,P_{n}$ is as follows:
$$P_1\subset P_2\subset\cdots\subset P_{n-1}\subset P_n.$$
The $P_{n}\setminus P_k$ stands for the set consisting of non-$k$-producible quantum states.

A rational $(k+1)$-partite entanglement measure $E$ should fulfill the following requirements:

(P1) $E(\rho)=0$ for any $\rho\in P_k$ and $E(\rho)>0$ for any $\rho\in P_{n}\setminus P_k$.

(P2) $E(\rho)$ satisfies the invariance under local unitary transformations.

(P3) Monotonicity, $E(\rho)$ does not increase under local operation and classical communication (LOCC), namely, $E(\Lambda_{\rm LOCC}(\rho))\leq E(\rho)$.

In addition, a host of entanglement measures may also possess the conditions:

(P4) Strong monotonicity, $E(\rho)\geq\sum_ip_iE(\sigma_i)$, where $\sigma_i$ is obtained with probability $p_i$ by $\Lambda_{\rm LOCC}$ performing on $\rho$.

(P5) Convexity, $E(\sum_ip_i\rho_i)\leq\sum_ip_iE(\rho_i)$.

(P6) Subadditivity, $E(\rho\otimes\sigma)\leq E(\rho)+E(\sigma)$.

Multipartite quantum states, which have multiple different entanglement ways, are highly complicated. Although there are a number of entanglement measures to characterize and quantify entanglement of states, they are far from complete. In this paper, we will present a series of general entanglement measures which can be employed to answer the question ``how many particles are entangled for a quantum state".

\section{The $(k+1)$-partite entanglement measures}\label{III}
The concept of $k$-producibility is of exceeding significance in quantum nonlocality \cite{31,32,33,34} and quantum metrology \cite{35}. The key of state characterization is to find rational entanglement measures. Consequently, we will present two classes of general $n$-partite entanglement measures in this section, termed $q$-$(k+1)$-PE concurrence and $\alpha$-$(k+1)$-PE concurrence, respectively, which are defined from the perspective of $(k+1)$-partite entanglement.

{\bf Definition 1}. For any $n$-partite pure state $|\phi\rangle$, the $q$-$(k+1)$-PE concurrence is defined as
\begin{equation}\label{1}
\begin{array}{rl}
E_{q-k}(|\phi\rangle)=\min\limits_{A}\sqrt{\frac{\sum_{t=1}^{m}[1-{\rm Tr}(\rho_{A_t}^{q})]}{m}},
\end{array}
\end{equation}
and the $\alpha$-$(k+1)$-PE concurrence is given by
\begin{equation}
\begin{array}{rl}\label{2}
E_{\alpha-k}(|\phi\rangle)=\min\limits_{A}\sqrt{\frac{\sum_{t=1}^{m}[{\rm Tr}(\rho_{A_t}^{\alpha})-1]}{m}}.
\end{array}
\end{equation}
Here $\rho_{A_t}={\rm Tr}_{{\overline A}_t}(|\phi\rangle\langle\phi|)$, ${\overline A}_t$ is the complement of subsystem $A_t$, $A=\{A_1|A_2|\cdots|A_m\}$ is the set satisfying the conditions (a), (b), and $|A_t|\leq k$ for any $t$, and the minimum runs over all possible elements of set $A$.

Note that the $k$ in Eqs. (\ref{1}) and (\ref{2}) associates with the maximum of $|A_t|$ under all possible partitions, $|A_t|\leq k$,  which is a completely different concept from $k$ in $k$-separable \cite{3,4,18}. Moreover, $m$ in Eqs. (\ref{1}) and (\ref{2}) is varied with regard to distinct partitions. So it can be seen that $q$-$k'$-PE concurrence differs from $q$-$k$-ME concurrence $C_{q-k}$ \cite{18}, and $\alpha$-$k'$-PE concurrence from $\alpha$-$k$-ME concurrence $C_{\alpha-k}$ \cite{18}, where $k$ and $k'$ goes from 2 to $n$. In particular, we obtain that the relations $E_{q-1}(|\phi\rangle)=\sqrt{C_{q-n}(|\phi\rangle)}$ and $E_{\alpha-1}(|\phi\rangle)=\sqrt{C_{\alpha-n}(|\phi\rangle)}$ holds for any pure state $|\phi\rangle$. When $q=2$, the $q$-$1$-PE concurrence is accorded with  $n$-ME concurrence \cite{3}. When $n=2$, $E_{q-1}(|\phi\rangle)=\sqrt{C_q(|\phi\rangle)}$, $E_{\alpha-1}(|\phi\rangle)=\sqrt{C_{\alpha}(|\phi\rangle)}$, so Eqs. (\ref{1}) and (\ref{2}) can be regarded as generalizations of $q$-concurrence $C_q$ \cite{5} and $\alpha$-concurrence $C_{\alpha}$ \cite{16}, respectively. When $n=2$ and $q=2$, $\sqrt{2}E_{2-1}(|\phi\rangle)=C(|\phi\rangle)$, thus concurrence $C$ is a special case of $q$-$(k+1)$-PE concurrence.

Let $|\phi\rangle$ be any pure state, we take $q$ as $a$ and $b$, respectively, and $1<a\leq b$, then $E_{a-k}(|\phi\rangle)\leq E_{b-k}(|\phi\rangle)$, that is to say, $E_{q-k}(|\phi\rangle)$ is a monotonically increasing function with respect to $q$. As $q$ proceeds to positive infinity, $E_{q-k}(|\phi\rangle)$ tends to one. When $\alpha$ is taken as $c$ and $d$, and $0\leq c\leq d<1$, $E_{c-k}(|\phi\rangle)\geq E_{d-k}(|\phi\rangle)$, which means that $E_{\alpha-k}(|\phi\rangle)$ is a monotonically decreasing function with respect to $\alpha$. Specially, when $\alpha=0$, one has $E_{\alpha-k}(|\phi\rangle)=\min\limits_{A}\sqrt{{[\sum_{t=1}^{m}(\gamma_{A_t}-1)]}/{m}}$, where $\gamma_{A_t}$ denotes the rank of reduced density matrix $\rho_{A_t}$ of $|\phi\rangle$.

In the following we generalize Definition 1 by convex roof extension, making them applicable to arbitrary mixed states $\rho$ of $n$-particles. The $q$-$(k+1)$-PE concurrence is defined as
\begin{equation}\label{3}
\begin{array}{rl}
E_{q-k}(\rho)=\inf\limits_{\{p_i,|\phi_i\rangle\}}\sum_{i}p_iE_{q-k}(|\phi_i\rangle),
\end{array}
\end{equation}
and the $\alpha$-$(k+1)$-PE concurrence is given by
\begin{equation}\label{11}
\begin{array}{rl}
E_{\alpha-k}(\rho)=\inf\limits_{\{p_i,|\phi_i\rangle\}}\sum_{i}p_iE_{\alpha-k}(|\phi_i\rangle).
\end{array}
\end{equation}
Here the infimum is taken over all feasible ensemble decompositions of $\rho$.

Next we will show in detail that these two types of quantification fashions comply with all the fundamental properties (P1) to (P6), rendering them intriguing entanglement measures to be considered in quantum information theory.

{\bf Theorem 1}. Both the $q$-$(k+1)$-PE concurrence and $\alpha$-$(k+1)$-PE concurrence serve as reasonable multipartite entanglement measures.

{\bf Proof}. It is obvious that $q$-$(k+1)$-PE concurrence and $\alpha$-$(k+1)$-PE concurrence possess the desired properties (P2) and (P5).

The following we verify rigorously that $E_{q-k}(\rho)$ satisfies the properties (P1), (P3), (P4), and (P6).

(P1) It is easy to derive $E_{q-k}(\rho)\geq0$ for any quantum state. For any pure state $|\phi\rangle\in P_k$, there is always some partition such that $|\phi\rangle=\otimes_{t=1}^{m}|\phi_t\rangle_{A_t}$ and $|A_{t}|\leq k$, so we can get $\rho_{A_t}=|\phi_t\rangle_{A_t}\langle\phi_t|$, and further obtain $1-{\rm Tr}(\rho_{A_t}^q)=0$ for any $t$, which means $E_{q-k}(|\phi\rangle)=0$. For any mixed state $\rho=\sum_ip_i|\phi_i\rangle\langle\phi_i|\in P_k$, by definition, $E_{q-k}(\rho)\leq\sum_ip_iE_{q-k}(|\phi_i\rangle)=0$. To sum up, $E_{q-k}(\rho)=0$ holds for any $\rho\in P_k$.

The calculation of $q$-$(k+1)$-PE concurrence requires to ensure the cardinality $|A_t|\leq k$ of each subsystem under any partition. For any pure state $|\phi\rangle\in P_n\backslash P_k$, it must be the case that there is certain subsystem $A_t$ correlated with ${\overline A}_t$ for each partition, so $1-{\rm Tr}(\rho_{A_t}^q)>0$, which implies $E_{q-k}(|\phi\rangle)>0$. For any mixed state $\rho\in P_n\backslash P_k$, it cannot be expressed as the convex combination of $k$-producible pure states, thus we have $E_{q-k}(\rho)>0$.

(P3) We will show that $q$-$(k+1)$-PE concurrence satisfies the property of entanglement monotone. For any pure state $|\phi\rangle$, if $\Lambda_{\rm LOCC}(|\phi\rangle)$ is a pure state, one has
\begin{equation*}
\begin{array}{rl}
&E_{q-k}(\Lambda_{\rm LOCC}(|\phi\rangle))\\
=&\min\limits_A\sqrt{\frac{\sum_{t=1}^{m}C_{q A_t|{\overline A}_t}(\Lambda_{\rm LOCC}(|\phi\rangle))}{m}}\\
\leq&\min\limits_A\sqrt{\frac{\sum_{t=1}^{m}C_{q A_t|{\overline A}_t}(|\phi\rangle)}{m}}\\
=&E_{q-k}(|\phi\rangle).\\
\end{array}
\end{equation*}
Here the inequality holds because $C_{q A_t|{\overline A}_t}(|\phi\rangle)=1-{\rm Tr}(\rho_{A_t}^q)$ is non-increasing under LOCC \cite{5}. If $\Lambda_{\rm LOCC}(|\phi\rangle)$ is a mixed state with ensemble decomposition $\{p_i, |\phi_i\rangle\}$ , we derive
\begin{equation*}
\begin{array}{rl}
&E_{q-k}(\Lambda_{\rm LOCC}(|\phi\rangle))\\
\leq&\sum_ip_iE_{q-k}(|\phi_i\rangle)\\
=&\sum_ip_i\min\limits_A\sqrt{\frac{\sum_{t=1}^{m}C_{q A_t|{\overline A}_t}(|\phi_i\rangle)}{m}}\\
\leq&\min\limits_A\sum_ip_i\sqrt{\frac{\sum_{t=1}^{m}C_{q A_t|{\overline A}_t}(|\phi_i\rangle)}{m}}\\
\leq&\min\limits_A\sqrt{\frac{\sum_{t=1}^{m}\sum_ip_i C_{q A_t|{\overline A}_t}(|\phi_i\rangle)}{m}}\\
\leq&\min\limits_A\sqrt{\frac{\sum_{t=1}^{m}C_{q A_t|{\overline A}_t}(|\phi\rangle)}{m}}\\
=&E_{q-k}(|\phi\rangle),\\
\end{array}
\end{equation*}
where the first inequality is based on the definition of $E_{q-k}(\rho)$, the third inequality is due to the concavity of $y=\sqrt{x}$, and the last inequality is true because $C_{q}(\rho)$ satisfies strong monotonicity.

Given a quantum state $\rho$, fancy that $\{p_i,|\phi_i\rangle\}$ is the optimal ensemble decomposition of $E_{q-k}(\rho)$.  Exploiting the definition of $E_{q-k}(\rho)$ and the monotonicity of $E_{q-k}$ for arbitrary pure states, we obtain
\begin{equation*}
\begin{array}{rl}
&E_{q-k}(\Lambda_{\rm LOCC}(\rho))\\
\leq&\sum_{i}p_iE_{q-k}(\Lambda_{\rm LOCC}(|\phi_i\rangle))\\
\leq&\sum_{i}p_iE_{q-k}(|\phi_i\rangle)\\
=&E_{q-k}(\rho).
\end{array}
\end{equation*}

Therefore $E_{q-k}(\rho)$ does not increase for any quantum state under LOCC.

(P4) Because $q$-concurrence meets strong monotonicity for any bipartite quantum state \cite{5}, the relation $C_{q A_{t}|{\overline A}_{t}}(\rho)\geq\sum_i p_i C_{qA_{t}|{\overline A}_{t}}(\sigma_{i})$ is true, where $\{p_i,\sigma_i\}$ is gotten by $\Lambda_{\rm LOCC}$ working on $\rho$. When $\rho=|\phi\rangle\langle\phi|$ is a pure state, $\sigma_i=K_i|\phi\rangle\langle\phi|K_i^\dagger$ is also a pure state yielded with the probability $p_i$, $\sum_iK_i^\dagger K_i=\mathbb{I}$, $\mathbb{I}$ is a unit operator. Supposed that $A_1|A_2|\cdots|A_m$ is the partition of $\rho$ such that $E_{q-k}(\rho)=\sqrt{\frac{[\sum_{t=1}^m(1-{\rm Tr}\rho_{A_t}^q)]}{{m}}}$, one sees
\begin{equation*}
\begin{array}{rl}
E_{q-k}(\rho)&=\sqrt{\frac{\sum_{t=1}^m(1-{\rm Tr}\rho_{A_t}^q)}{m}}\\
&=\sqrt{\frac{\sum_{t=1}^mC_{qA_{t}|{\overline A}_{t}}(|\phi\rangle)}{m}}\\
&\geq\sqrt{\frac{\sum_{t=1}^m\sum_{i}p_iC_{qA_{t}|{\overline A}_{t}}(\sigma_i)}{m}}\\
&\geq\sum_{i}p_i\sqrt{\frac{\sum_{t=1}^mC_{qA_{t}|{\overline A}_{t}}(\sigma_i)}{m}}\\
&\geq\sum_{i}p_iE_{q-k}(\sigma_i).\\
\end{array}
\end{equation*}
Here the second inequality is because $y=\sqrt{x}$ is a concave function, the last inequality is true following from Eq. (\ref{3}).

According to the definition of $E_{q-k}(\rho)$ given in Eq. (\ref{3}) and the fact that $E_{q-k}$ fulfills strong monotonicity for arbitrary pure states, the conclusion that $E_{q-k}(\rho)$ satisfies strong monotonicity for any mixed state can be drawn.

(P6) For arbitrary two pure states $\rho=|\phi\rangle\langle\phi|$ and $\sigma=|\psi\rangle\langle\psi|$, assumed that $A_1|A_2|\cdots|A_{m_1}$ and $B_1|B_2|\cdots|B_{m_2}$ are respectively optimal partitions of $\rho$ and $\sigma$ such that $E_{q-k}(\rho)=\sqrt{\frac{\sum_{t=1}^{m_1}[1-{\rm Tr}(\rho_{A_t}^q)]}{{m_1}}}$, $E_{q-k}(\sigma)=\sqrt{\frac{\sum_{t=1}^{{m_2}}[1-{\rm Tr}(\sigma_{B_t}^q)]}{{m_2}}}$, then we have
\begin{equation}\label{7}
\begin{array}{rl}
E_{q-k}(\rho\otimes\sigma)&\leq\sqrt{\frac{\sum_{t=1}^{m_1}[1-{\rm Tr}(\rho_{A_t}^q)]+\sum_{t=1}^{m_2}[1-{\rm Tr}(\sigma_{B_t}^q)]}{m_1+m_2}}\\
&\leq\sqrt{\frac{\sum_{t=1}^{m_1}[1-{\rm Tr}(\rho_{A_t}^q)]}{m_1}+\frac{\sum_{t=1}^{m_2}[1-{\rm Tr}(\sigma_{B_t}^q)]}{m_2}}\\
&\leq\sqrt{\frac{\sum_{t=1}^{m_1}[1-{\rm Tr}(\rho_{A_t}^q)]}{m_1}}+\sqrt{\frac{\sum_{t=1}^{m_2}[1-{\rm Tr}(\sigma_{B_t}^q)]}{m_2}}\\
&=E_{q-k}(\rho)+E_{q-k}(\sigma).\\
\end{array}
\end{equation}

Using the convexity of $E_{q-k}(\rho)$ and the relation presented in inequality (\ref{7}), we can verify that $E_{q-k}(\rho)$ also fulfills subadditivity for any mixed state.

Adopting analogous procedures, we can directly testify that $\alpha$-$(k+1)$-PE concurrence also obeys the requirements (P1), (P3), (P4), and (P6). $\hfill\blacksquare$

Therefore, the $q$-$(k+1)$-PE concurrence and $\alpha$-$(k+1)$-PE concurrence satisfy all the properties to be an entanglement measure, which means that these two types of methods can be used to explicitly detect whether a quantum state is $k$-partite entangled. If $E_{p-k}(\rho)=0$, then $\rho$ is a $k$-producible state; if $E_{p-(k-1)}(\rho)>0$ and $E_{p-k}(\rho)=0$, then $\rho$ is a genuinely $k$-producible state, where $p=q~{\rm or}~\alpha$. In particular, if $E_{p-1}(\rho)=0$, then $\rho$ is a fully separable state; if $E_{p-(n-1)}(\rho)>0$, then $\rho$ is a genuinely entangled state. Moreover, these characteristics favor them as potential quantum resources.

The $\alpha$-$k$-ME concurrence $(2\leq k\leq n)$ does not satisfy subadditivity, whereas $\alpha$-$k$-PE concurrence $(2\leq k\leq n)$ does, which also shows that $\alpha$-$k$-PE concurrence and $\alpha$-$k$-ME concurrence are two completely different methods of quantification.

In Ref. \cite{4}, Gao $et~al$. claimed that every entanglement measure $E$ should obey the demand that the maximum of $E(\rho_U^{\rm PI})$ under any local unitary transformation $U$ is a lower bound of $E(\rho)$, where $\rho^{\rm PI}$ is the permutationally invariant (PI) part of $\rho$. For any individual raised measure, however, one must demonstrate that it is indeed met. We will show that $q$-$(k+1)$-PE concurrence and $\alpha$-$(k+1)$-PE concurrence fulfill the requirement $E(\rho)\geq\max_U E(\rho_U^{\rm PI})$.

{\bf Theorem 2}. For an arbitrary $n$-partite quantum state $\rho$, the lower bound of $q$-$(k+1)$-PE concurrence of $\rho$ is given by means of the maximum of $q$-$(k+1)$-PE concurrence of PI part of $\rho$, that is,
\begin{equation}\label{4}
\begin{array}{rl}
E_{q-k}(\rho)\geq\max\limits_U E_{q-k}({\rho_{U}^{\rm PI}}).\\
\end{array}
\end{equation}
For the $\alpha$-$(k+1)$-PE concurrence, an analogous relation is as follows:
\begin{equation}\label{5}
\begin{array}{rl}
E_{\alpha-k}(\rho)\geq\max\limits_U E_{\alpha-k}({\rho_{U}^{\rm PI}}).\\
\end{array}
\end{equation}
Here $U$ is any locally unitary transformation.

{\bf Proof}. If $A_1|A_2|\cdots|A_m$ is a partition of $n$ subsystems, then it is easy to see that $\Pi_j(A_1)|\Pi_j(A_2)|\cdots|\Pi_j(A_m)$ is also a partition of $n$ subsystems, where $|A_t|\leq k$ and $t$ ranges from 1 to $m$. Supposed that $|\phi\rangle$ is an any pure state, then $\Pi_j(|\phi\rangle)$ is also a pure state and the following relation holds,
\begin{equation}\label{6}
\begin{array}{rl}
E_{q-k}(|\phi\rangle)=E_{q-k}(\Pi_j(|\phi\rangle)),\\
\end{array}
\end{equation}
where $\Pi_j$ is any $n$-element permutation.

For any $n$-partite pure state $|\phi\rangle$, the PI part of $\rho=|\phi\rangle\langle\phi|$ can be represented as $\rho^{\rm PI}=\frac{1}{n!}\sum_{i=1}^{n!}\Pi_j|\phi\rangle\langle\phi|\Pi_j^{\dagger}$, one can get
\begin{equation*}
\begin{array}{rl}
E_{q-k}(\rho^{\rm PI})&\leq \frac{1}{n!}\sum_{j=1}^{n!}E_{q-k}(\Pi_j|\phi\rangle)\\
&=\frac{1}{n!}\sum_{j=1}^{n!}E_{q-k}(|\phi\rangle)\\
&=E_{q-k}(|\phi\rangle).
\end{array}
\end{equation*}
Here the inequality is owing to the convexity of $E_{q-k}(\rho)$, the first equality is derived from the above formula (\ref{6}).

Let $\rho=\sum_ip_i|\phi_i\rangle\langle\phi_i|$ be an $n$-partite mixed state, and $\rho^{\rm PI}=\sum_ip_i(|\phi_i\rangle\langle\phi_i|)^{\rm PI}$ \cite{4}. By utilizing the Eq. (\ref{6}) and the convexity of $E_{q-k}(\rho)$, we derive
\begin{equation}\label{8}
\begin{array}{rl}
E_{q-k}(\rho)\geq E_{q-k}(\rho^{\rm PI}).\\
\end{array}
\end{equation}

Owing to the fact that the PI part of a quantum state relies on the choice of bases \cite{4}, and the inequality (\ref{8}) is true under any locally unitary transformation $U$, so we can directly see
\begin{equation*}
\begin{array}{rl}
E_{q-k}(\rho)\geq \max\limits_U E_{q-k}(\rho_U^{\rm PI}).\\
\end{array}
\end{equation*}

The inequality (\ref{5}) can be testify by utilizing analogical methods. $\hfill\blacksquare$

Therefore, these parametrized $(k+1)$-partite entanglement measures $E_{q-k}(\rho)$ and $E_{\alpha-k}(\rho)$ comply with the requirement proposed in Ref. \cite{4}. The right sides of inequalities (\ref{4}) and (\ref{5}) are the strong lower bounds of $E_{q-k}(\rho)$ and $E_{\alpha-k}(\rho)$, respectively. This also illustrates that if $\rho^{\rm PI}$ is entangled with $(k+1)$-partite, then so is the original state $\rho$ \cite{30}.

\section{The $(k+1)$-partite entanglement measures based on geometric mean}\label{IV}

Apart from taking the minimum of entanglement under all partitions in set $A$, as shown in Eqs. (\ref{1}) and (\ref{2}), we will provide two classes of alternative entanglement measures for multipartite quantum states based on the geometric mean in this section.

{\bf Definition 2}. For any $n$-partite pure state $|\phi\rangle$, the $q$-$(k+1)$-GPE concurrence is defined as
\begin{equation}\label{9}
\begin{array}{rl}
\mathcal{E}_{q-k}(|\phi\rangle)&=\bigg(\frac{\prod_{\alpha_i\in A}[\sum_{t=1}^{m_i}(1-{\rm Tr}\rho_{A_{t{\alpha_i}}}^q)]}{\prod_{i=1}^{s_k(n)}m_{i}}\bigg)^{\frac{1}{2s_k(n)}},
\end{array}
\end{equation}
and the $\alpha$-$(k+1)$-GPE concurrence is given by
\begin{equation}\label{10}
\begin{array}{rl}
\mathcal{E}_{\alpha-k}(|\phi\rangle)=\bigg(\frac{\prod_{\alpha_i\in A}[\sum_{t=1}^{m_i}({\rm Tr}\rho_{A_{t{\alpha_i}}}^{\alpha}-1)]}{\prod_{i=1}^{s_k(n)}m_{i}}\bigg)^{\frac{1}{2s_k(n)}}.\\
\end{array}
\end{equation}
Here $A=\{\alpha_i\}=\{A_{1\alpha_i}|A_{2\alpha_i}|\cdots|A_{m_i\alpha_i}\}$ is the set of all partitions satisfying the conditions (a), (b), and $|A_{t{\alpha_i}}|\leq k$, $s_k(n)$ stands for the cardinality of elements in the set $A$, $\rho_{A_{t{\alpha_i}}}$ is the reduced density operator with respect to subsystem ${A_{t{\alpha_i}}}$, and $m_i$ indicates the number of subsystems for partition $\alpha_i$. Several specific $s_k(n)$ are listed in Table \ref{tab:t1}.

Although the representation of set $A$ in Definition 1 is slightly different from that in Definition 2, they contain essentially the same elements. Moreover, it is worth mentioning that these two types of quantifications shown in Eqs. (\ref{9}) and (\ref{10}) are related to all entanglement values under partitions satisfying preconditions rather than relying on the minimum of entanglement.

The above definition can be extended to apply to arbitrary quantum states. For any $n$-partite mixed state $\rho$, we define the $q$-$(k+1)$-GPE concurrence as
\begin{equation}\label{12}
\begin{array}{rl}
\mathcal{E}_{q-k}(\rho)=\inf\limits_{\{p_i,|\phi_i\rangle\}}\sum_{i}p_i\mathcal{E}_{q-k}(|\phi_i\rangle),
\end{array}
\end{equation}
and the $\alpha$-$(k+1)$-GPE concurrence as
\begin{equation}\label{13}
\begin{array}{rl}
\mathcal{E}_{\alpha-k}(\rho)=\inf\limits_{\{p_i,|\phi_i\rangle\}}\sum_{i}p_i\mathcal{E}_{\alpha-k}(|\phi_i\rangle).
\end{array}
\end{equation}
Here the infimum runs over all feasible pure state decompositions of $\rho$.

\begin{table}[htbp]
\centering
 \caption{\label{tab:test} The cardinalities of all feasible partitions satisfying $|A_{t{\alpha_i}}|\leq k$ are listed for $n$-partite quantum systems, where $n$ takes 3,~4,~5,~6,~7,~8, respectively, and $k$ goes from 1 to $n-1$.}
 \label{tab:t1}
 \begin{tabular}{cccccccc}
  \hline\hline

 ~~&~~~$\emph{k}$=1~~  & $~~\emph{k}$=2~~ & $~~\emph{k}$=3~~  & $~~\emph{k}$=4~~ & $~~\emph{k}$=5~~ & $~~\emph{k}$=6~~ & $\emph{k}$=7\\
 \hline
 \vspace{0.3em}
 $n=3$ & 1 & 4 & - & - & - & - & -\\
 \vspace{0.3em}
 $n=4$ & 1 & 10 & 14 & - & - & - & -\\
 \vspace{0.3em}
 $n=5$ & 1 & 26 & 46 & 51 & - & - & -\\
 \vspace{0.3em}
 $n=6$ & 1 & 76 & 166 & 196 & 202 & - & -\\
 \vspace{0.3em}
 $n=7$ & 1 & 232 & 652 & 827 & 869 & 876 & -\\
 \vspace{0.3em}
 $n=8$ & 1 & 764 & 2780 & 3795 & 4075 & 4131 & 4139\\
  \hline\hline
 \end{tabular}
\end{table}

It is obvious that $q$-$k'$-GPE concurrence differs from $q$-$k$-GM concurrence $\mathcal{G}_{q-k}(\rho)$ \cite{18} and $\alpha$-$k'$-GPE concurrence from $\alpha$-$k$-GM concurrence $\mathcal{G}_{\alpha-k}(\rho)$ \cite{18}, where $2\leq k,~k'\leq n$.  However, for the special case, $n$-partition, we observe that the relations $\sqrt{2}\mathcal{E}_{q-1}(\rho)=\mathcal{G}_{q-n}(\rho)$ and $\sqrt{2}\mathcal{E}_{\alpha-1}(\rho)=\mathcal{G}_{\alpha-n}(\rho)$ are valid for arbitrary quantum states $\rho$.

The following we will show that $\mathcal{E}_{q-k}(\rho)$ and $\mathcal{E}_{\alpha-k}(\rho)$ satisfy the necessary conditions (P1) to (P5).

{\bf Theorem 3}. The $q$-$(k+1)$-GPE concurrence and $\alpha$-$(k+1)$-GPE concurrence are proper $(k+1)$-partite entanglement measures.

{\bf Proof}. Adopting a similar idea as in Theorem 1, we obtain that the property (P1) holds. And it is evident that the properties (P2) and (P5) are valid.

Next we verify that the $q$-$(k+1)$-GPE concurrence obeys monotonicity. Since $q$-concurrence does not increase under LOCC \cite{5}, for any pure state $|\phi\rangle$ of $n$-particles, we only need to prove that $\mathcal{E}_{q-k}(|\phi\rangle)$ is an increasing function of $C_{qA_{t{\alpha_j}}|\overline{A}_{t{\alpha_j}}}(|\phi\rangle)$, one sees
\begin{equation*}
\begin{array}{rl}
&\frac{\partial\mathcal{E}_{q-k}(|\phi\rangle)}{\partial C_{qA_{t\alpha_j}|{\overline A}_{t\alpha_j}}(|\phi\rangle)}\\
=&\frac{\prod\limits_{A\backslash \{\alpha_j\}}\Big[{\sum\limits_{t=1}^{m_i}C_{q{{A_{t{\alpha_i}}}|{{\overline A}_{t{\alpha_i}}}}}(|\phi\rangle)}\Big]}{2s_k(n)\Big(\prod\limits_{i=1}^{s_k(n)}m_i\Big)^{\frac{1}{2s_k(n)}}\Big(\prod\limits_{\alpha_i\in A}\big[{\sum\limits_{t=1}^{m_i}C_{q{{A_{t{\alpha_i}}}|{{\overline A}_{t{\alpha_i}}}}}(|\phi\rangle)}\big]\Big)^{\frac{2s_k(n)-1}{2s_k(n)}}}\\
\geq&0,
\end{array}
\end{equation*}
where $C_{qA_{t{\alpha_j}}|\overline{A}_{t{\alpha_j}}}(|\phi\rangle)=1-{\rm Tr}(\rho_{A_{t{\alpha_j}}}^q)$, $j$ ranges from 1 to $s_k(n)$. So the $\mathcal{E}_{q-k}(|\phi\rangle)$ is entangled monotone.

Following from the definition of $\mathcal{E}_{q-k}(\rho)$ and the monotonicity of $\mathcal{E}_{q-k}(|\phi\rangle)$, we get easily that $\mathcal{E}_{q-k}(\rho)$ is non-increasing for any mixed state $\rho$ under LOCC.

At last, we prove that the $q$-$(k+1)$-GPE concurrence satisfies strong monotonicity. Because $q$-concurrence meets the relation $C_q(\rho)\geq\sum_jp_jC_q(\sigma_j)$ \cite{5}, where $\{p_j,\sigma_j\}$ is an ensemble yielded by $\Lambda_{\rm LOCC}$ acting on $\rho$. Let us consider any $n$-partite pure state $|\phi\rangle$, $\sigma_j=K_j|\phi\rangle\langle\phi|K_j^\dagger$, where $\sum_jK_j^\dagger K_j=\mathbb{I}$, and the following relation can be derived,

\begin{equation*}
\begin{array}{rl}
\mathcal{E}_{q-k}(|\phi\rangle)&=\bigg(\frac{\prod_{\alpha_i\in A}[\sum_{t=1}^{m_i}C_{qA_{t\alpha_i}|\overline{A}_{t\alpha_i}}(|\phi\rangle)]}{\prod_{i=1}^{s_k(n)}m_i}\bigg)^{\frac{1}{2s_k(n)}}\\
&\geq\bigg(\frac{\prod_{\alpha_i\in A}\sqrt{\sum_{t=1}^{m_i}\sum_jp_jC_{qA_{t\alpha_i}|\overline{A}_{t\alpha_i}}(\sigma_j)}}{\prod_{i=1}^{s_k(n)}\sqrt{m_i}}\bigg)^{\frac{1}{s_k(n)}}\\
&\geq\bigg(\frac{\prod_{\alpha_i\in A}[\sum_jp_j\sqrt{\sum_{t=1}^{m_i}C_{qA_{t\alpha_i}|\overline{A}_{t\alpha_i}}(\sigma_j)}]}{\prod_{i=1}^{s_k(n)}\sqrt{m_i}}\bigg)^{\frac{1}{s_k(n)}}\\
&\geq\sum_jp_j\bigg(\frac{\prod_{\alpha_i\in A}\sqrt{\sum_{t=1}^{m_i}C_{qA_{t\alpha_i}|\overline{A}_{t\alpha_i}}(\sigma_j)}}{\prod_{i=1}^{s_k(n)}\sqrt{m_i}}\bigg)^{\frac{1}{s_k(n)}}\\
&=\sum_jp_j\mathcal{E}_{q-k}(|\phi_j\rangle).\\
\end{array}
\end{equation*}
Here the second inequality is due to the concavity of $y=\sqrt{x}$, the third inequality follows from the fact that $y=({\prod_{i=1}^{n}x_i})^{\frac{1}{n}}$ is a concave function, and the last equality holds according to the Eq. (\ref{9}).

Based on the definition of $\mathcal{E}_{q-k}(\rho)$ and $\mathcal{E}_{q-k}$ fulfills strong monotonicity for any pure state, it is not difficult to verify that $\mathcal{E}_{q-k}$ also possesses the property (P4) for any mixed state.

Using similar approaches, we can show that $\alpha$-$(k+1)$-GPE concurrence also satisfies properties (P3) and (P4), i.e., monotonicity and strong monotonicity. $\hfill\blacksquare$

We see that these two classes of entanglement measures can also definitively detect whether a quantum state is $k$-producible, which plays an essential role in quantum metrology. If $\mathcal{E}_{p-k}(\rho)=0$, then $\rho$ is a $k$-producible state; if $\mathcal{E}_{p-(k-1)}(\rho)>0$ and $\mathcal{E}_{p-k}(\rho)=0$, then $\rho$ is a genuinely $k$-producible state, where $p=q~{\rm or}~\alpha$.

Unfortunately, the calculation of entanglement for mixed states is extremely difficult, which prompts us to give lower bounds for $q$-$(k+1)$-GPE concurrence and $\alpha$-$(k+1)$-GPE concurrence.

{\bf Theorem 4}. For an arbitrary $n$-partite quantum state $\rho$, the lower bound of $q$-$(k+1)$-GPE concurrence of $\rho$ is given by means of the maximum of $q$-$(k+1)$-GPE concurrence of PI part of $\rho$, namely,
\begin{equation}
\begin{array}{rl}
\mathcal{E}_{q-k}(\rho)\geq\max\limits_U \mathcal{E}_{q-k}({\rho_{U}^{\rm PI}}).\\
\end{array}
\end{equation}
For the $\alpha$-$(k+1)$-GPE concurrence, there is an analogous relation,
\begin{equation}
\begin{array}{rl}
\mathcal{E}_{\alpha-k}(\rho)\geq\max\limits_U \mathcal{E}_{\alpha-k}({\rho_{U}^{\rm PI}}).\\
\end{array}
\end{equation}
Here the maximum is taken under all locally unitary transformations $U$.

The proof of this theorem is similar to the proof of Theorem 2, for the sake of brevity, we will not explain it here. This conclusion illustrates that $q$-$(k+1)$-GPE concurrence and $\alpha$-$(k+1)$-GPE concurrence satisfy the requirement set forth in Ref. \cite{4}, which implies that the dimension of the considered space is greatly reduced when we judge the entangled structure of a quantum state.

We proceed to note that there is a relation between $q$-$(k+1)$-PE concurrence and $q$-$(k+1)$-GPE concurrence, which is provided by the following theorem.

{\bf Theorem 5}. The $q$-$(k+1)$-GPE concurrence is lower bounded by the $q$-$(k+1)$-PE concurrence.

{\bf Proof}. For any pure state $|\phi\rangle$ with $n$-particles, we can get obviously
\begin{equation}
\begin{array}{rl}
\mathcal{E}_{q-k}(|\phi\rangle)\geq E_{q-k}(|\phi\rangle).\\
\end{array}
\end{equation}

For any mixed state $\rho$, let $\{p_i, |\phi_i\rangle\}$ be the optimal pure decomposition of $\mathcal{E}_{q-k}(\rho)$, then one sees
\begin{equation}
\begin{array}{rl}
\mathcal{E}_{q-k}(\rho)&=\sum_ip_i\mathcal{E}_{q-k}(|\phi_i\rangle)\\
&\geq\sum_ip_iE_{q-k}(|\phi_i\rangle)\\
&\geq E_{q-k}(\rho).\\
\end{array}
\end{equation}

The proof is concluded. $\hfill\blacksquare$

Analogously, the conclusion that the $\alpha$-$(k+1)$-PE concurrence is a lower bound of $\alpha$-$(k+1)$-GPE concurrence can be drawn directly.

\section{Comparison among entanglement measures}\label{V}
We present a series of entanglement measures from different perspectives, where one main line is to take the minimum of the entanglement, and the other main line is based on the geometric mean of the entanglement. They are $(k+1)$-partite entanglement measures with parameters, which can exhibit more properties of multipartite entangled states. Some of the things they have in common are enumerated here, for example:

(i) These forms of quantification possess the properties (P1) to (P5), namely, entanglement monotone, strong monotone, convexity, being zero for the states in set $P_k$, and strictly positive for the states belonging to the set $P_n\setminus P_k$.

(ii) All of them satisfy the requirement that the entanglement $E$ of a quantum state $\rho$ is lower bounded by the entanglement of PI part of $\rho$, where $E$ represents $E_{q-k}$, $E_{\alpha-k}$, $\mathcal{E}_{q-k}$, and $\mathcal{E}_{\alpha-k}$, respectively.

(iii) When $k=1$, the $q$-2-PE concurrence is consistent with the $q$-2-GPE concurrence, the $\alpha$-2-PE concurrence is accorded with the $\alpha$-2-GPE concurrence.

However, they exist differences in some areas. The details are explained as follows,

(iv) The $q$-$(k+1)$-GPE concurrence can distinguish some states sometimes that $q$-$(k+1)$-PE concurrence fails. For instance, given the states $|\phi_1\rangle=\frac{1}{2}(|0000\rangle+|1011\rangle+|1101\rangle+|1111\rangle)$ and $|\phi_2\rangle=\frac{1}{2}(|0000\rangle+|1111\rangle+|1001\rangle+|1110\rangle)$, we can deduce $E_{2-2}(|\phi_1\rangle)=E_{2-2}(|\phi_2\rangle)={\frac{\sqrt 6}{4}}$, $\mathcal{E}_{2-2}(|\phi_1\rangle)=\frac{\sqrt[20]{160000}}{\sqrt{8}}\neq\mathcal{E}_{2-2}(|\phi_2\rangle)=\frac{\sqrt[20]{756940800}}{4}$. From the results, the $q$-$(k+1)$-PE concurrence of a quantum state seems simpler than the $q$-$(k+1)$-GPE concurrence of that state here, however, the $q$-$(k+1)$-PE concurrence depends on the minimum of entanglement under all partitions, where the partitions satisfy conditions (a), (b), and the number of particles per subsystem does not exceed $k$, this may cause that the measures offered in Definition 1 ignore the global entanglement distribution of quantum states.

(v) The $q$-$(k+1)$-(G)PE concurrence is a monotonically increasing function with respect to $q$, whereas $\alpha$-$(k+1)$-(G)PE concurrence is a monotonically decreasing function with respect to $\alpha$.

(vi) The $q$-$(k+1)$-GPE concurrence and $\alpha$-$(k+1)$-GPE concurrence are smooth when the measured quantum state is varied continuously, while $q$-$(k+1)$-PE concurrence and $\alpha$-$(k+1)$-PE concurrence may appear peaks.

(vii) The entanglement orders of these measures are different at times.

To make this more intuitive, let us explain (vi) and (vii) using the following example.

\begin{figure}[htbp]
\centering
\subfigure[]{\includegraphics[width=8cm,height=6cm]{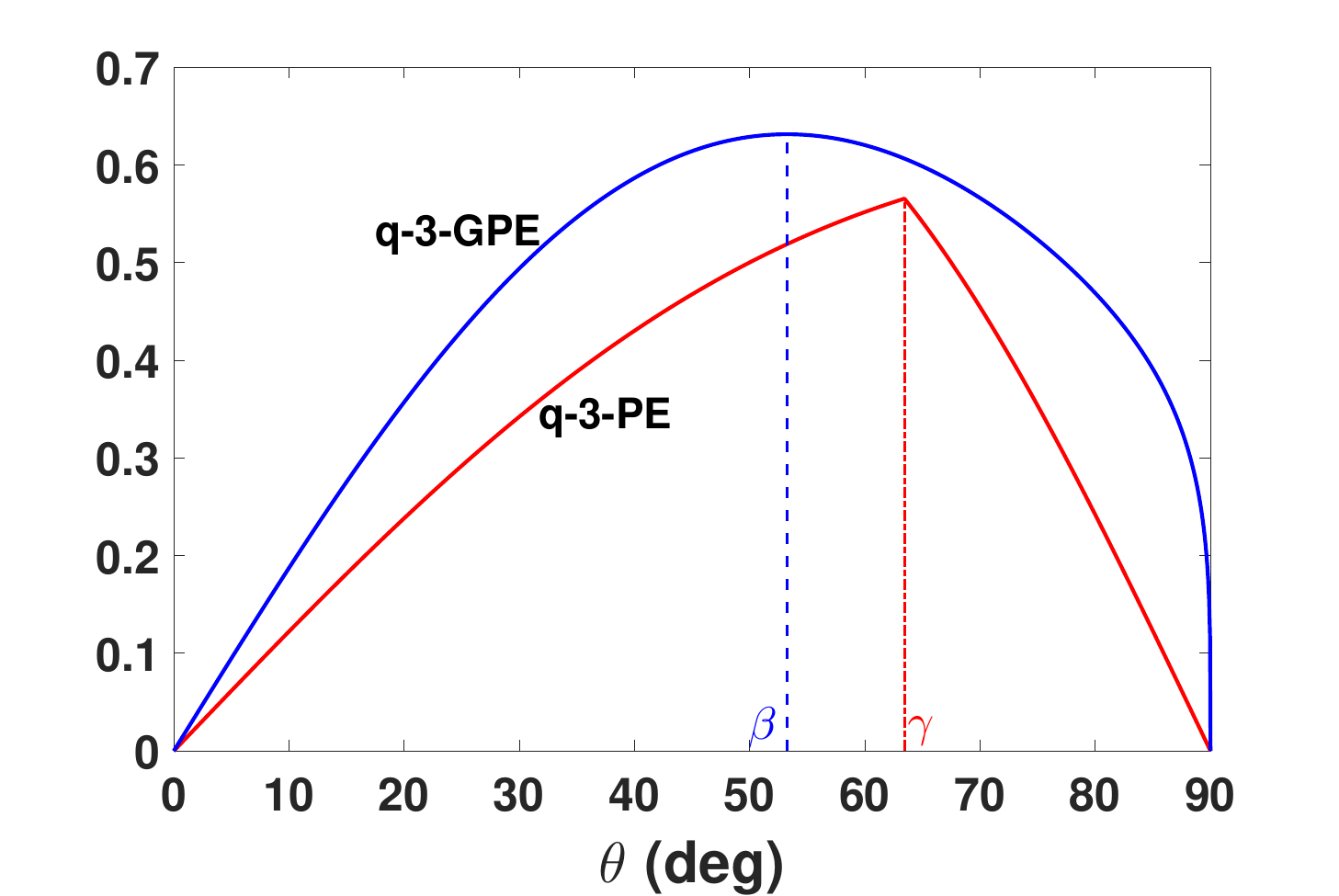}}
\quad
\subfigure[]{\includegraphics[width=8cm,height=6cm]{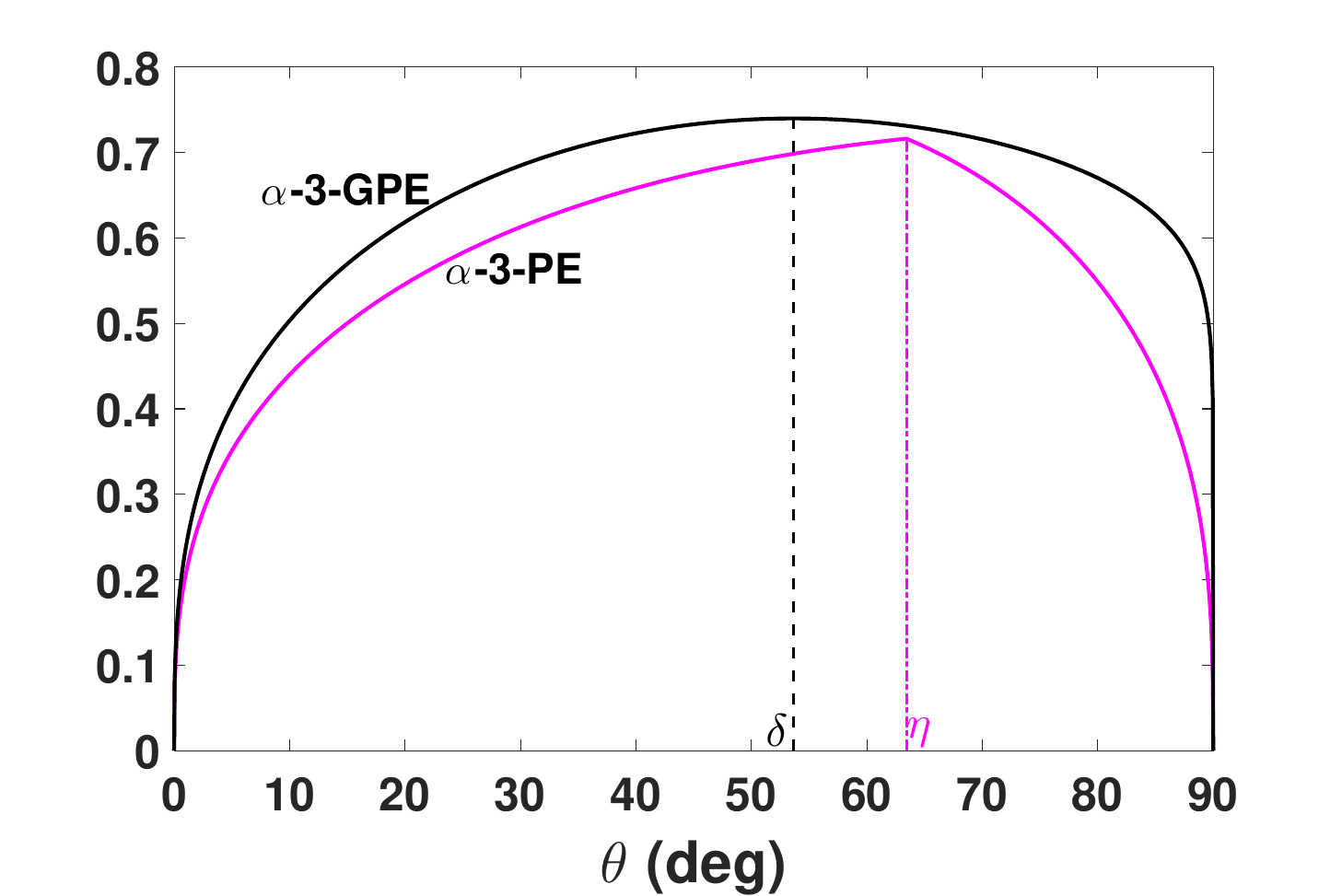}}
\caption{Set $k=2$. In (a), the red (lower) curve line stands for the $q$-3-PE concurrence of $|\phi\rangle$ and the blue (upper) curve line denotes $q$-3-GPE concurrence of $|\phi\rangle$ for $q=2$. In (b), the magenta (lower) curve line is the $\alpha$-3-PE concurrence of $|\phi\rangle$ and the black (upper) curve line expresses $\alpha$-3-GPE concurrence of $|\phi\rangle$ for $\alpha=\frac{1}{3}$.}\label{fig 1}
\end{figure}

{\bf Example 1}. Let $|\phi_\theta\rangle={\rm sin}\theta(\frac{1}{2}|010\rangle+\frac{\sqrt{3}}{2}|100\rangle)+{\rm cos}\theta|001\rangle$, $\theta\in[0,90^\circ]$, $k=2$, $q=2$, we evaluate $E_{2-2}(|\phi_\theta\rangle)$ and $\mathcal{E}_{2-2}(|\phi_\theta\rangle)$ and plot them in Fig. \ref{fig 1}(a). Then we observe the entanglement order of these two measures is different when $\theta$ belongs to the interval $(\beta,\gamma)$, i.e., when $\theta_1>\theta_2\in(\beta,\gamma)$, there is $E_{2-2}(|\phi_{\theta_1}\rangle)>E_{2-2}(|\phi_{\theta_2}\rangle)$, while $\mathcal{E}_{2-2}(|\phi_{\theta_1}\rangle)<\mathcal{E}_{2-2}(|\phi_{\theta_2}\rangle)$. In addition, $E_{2-2}(|\phi_\theta\rangle)$ appears a peak when $\theta=\gamma$ due to the minimization procedure. The similar conclusions can be obtained for $\alpha=\frac{1}{3}$, as shown in Fig. \ref{fig 1}(b).

These differences, which may be more conducive to our overall understanding of the characteristics of multipartite states, have important implications.

\section{The quantifications of non-genuinely strong $k$-producible states}\label{VI}
In the previous two sections we provided a series of entanglement measures that facilitate the determination of whether a quantum state is $k$-producible, and in this section several quantitative methods that help determine whether a multipartite pure state is genuinely $k$-producible will be presented initially.

A pure quantum state $|\phi\rangle$ is genuinely $k$-producible (or genuinely $k$-partite entanglement) \cite{2} if it is not $(k-1)$-producible, i.e., $\rho\in P_k\backslash P_{k-1}$. Therefore, we require to consider the partitions $A_1|A_2|\cdots|A_m$ to those that not only satisfy $|A_t|\leq k$, but also have at least one $A_t$ meeting $|A_t|=k$.

Now, we adjust ``$|A_t|\leq k$ for any $t$'' in Definition 1 to ``$|A_t|\leq k$ for any $t$ and at least one $A_t$ satisfying $|A_t|=k$'' ({\it labelled condition} (c)), and enforce ``$|A_{t{\alpha_i}}|\leq k$'' in Definition 2 to ``$|A_{t{\alpha_i}}|\leq k$ and at least one $A_{t{\alpha_i}}$ satisfying $|A_{t{\alpha_i}}|=k$'' ({\it marked condition} (c')), then the new quantification forms are as follows.

{\bf Definition 3}. Suppose that each partition for $n$-partite pure state $|\phi\rangle$ satisfies conditions (a), (b), and (c), then any bipartition is performed for the parts containing $k$ subsystems, we define the genuine $p$-$(k+1)$-PE ($p\geq0~{\rm and}~p\neq1$) is
\begin{equation}
\begin{array}{rl}
E'_{p-k}(|\phi\rangle)=a
\end{array}
\end{equation}
if there is a bipartition such that $A_{t}^{k_1}$ and $A_{t}^{k_2}$ are not related for any $A_{t}$ containing $k$-particles, and
\begin{equation}\label{14}
\begin{array}{rl}
E'_{p-k}(|\phi\rangle)=\min\limits_{A'}\sqrt{\frac{\sum_{t=1}^{m}|1-{\rm Tr}(\rho_{A_t}^{p})|}{m}}
\end{array}
\end{equation}
if for each partition $A_1|\cdots|A_m$ there is at least one $A_{t}$ with $k$-particles such that $A_{t}^{k_1}$ and $A_{t}^{k_2}$ are correlated ({\it marked condition} (d)). Here $a$ is a positive constant, $k_1+k_2=k$,  $A_{t}^{k_1}|A_{t}^{k_2}$ denotes any possible bipartition of $A_{t}$ including $k$-particles, $A'$ is the set consisted by the elements which satisfy the prerequisites (a), (b), (c), and (d), the minimum is obtained by traversing all elements of set $A'$.

Analogously, an alternative quantification methods are shown below.

\begin{table}[htbp]
\centering
 \caption{\label{tab:test} The cardinalities of all feasible partitions satisfying $|A_{t{\alpha_i}}|\leq k$ and at least one $|A_{t{\alpha_i}}|=k$  are listed for $n$-partite quantum systems, where $n$ takes 3,~4,~5,~6,~7,~8, respectively, and $k$ ranges from 1 to $n-1$.}
 \label{tab:t2}
 \begin{tabular}{cccccccc}
  \hline\hline

 ~~&~~~$\emph{k}$=1~~  & $~~\emph{k}$=2~~ & $~~\emph{k}$=3~~  & $~~\emph{k}$=4~~ & $~~\emph{k}$=5~~ & $~~\emph{k}$=6~~ & $\emph{k}$=7\\
 \hline
 \vspace{0.3em}
 $n=3$ & 1 & 3 & - & - & - & - & -\\
 \vspace{0.3em}
 $n=4$ & 1 & 9 & 4 & - & - & - & -\\
 \vspace{0.3em}
 $n=5$ & 1 & 25 & 20 & 5 & - & - & -\\
 \vspace{0.3em}
 $n=6$ & 1 & 75 & 90 & 30 & 6 & - & -\\
 \vspace{0.3em}
 $n=7$ & 1 & 231 & 420 & 175 & 42 & 7 & -\\
 \vspace{0.3em}
 $n=8$ & 1 & 763 & 2016 & 1015 & 280 & 56 & 8\\
  \hline\hline
 \end{tabular}
\end{table}

{\bf Definition 4}. Suppose that each partition for $n$-partite pure state $|\phi\rangle$ satisfies conditions (a), (b), and (c'), then any bipartition is performed for the parts containing $k$ subsystems, the genuine $p$-$(k+1)$-GPE ($p\geq0~{\rm and}~p\neq1$) is defined as
\begin{equation}
\begin{array}{rl}
\mathcal{E}'_{p-k}(|\phi\rangle)=b
\end{array}
\end{equation}
if there is a bipartition such that $A_{t\alpha_i}^{k_1}$ and $A_{t\alpha_i}^{k_2}$ are not related for all $A_{t\alpha_i}$ containing $k$-particles, and
\begin{equation}\label{16}
\begin{array}{rl}
\mathcal{E}'_{p-k}(|\phi\rangle)&=\bigg({\frac{\prod_{\alpha_i\in A'}[\sum_{t=1}^{m_{i}}|1-{\rm Tr}\rho_{A_{t{\alpha_i}}}^p|]}{\prod_{i=1}^{s'_k(n)}m_{i}}}\bigg)^{\frac{1}{2s_k(A')}},
\end{array}
\end{equation}
if for each partition $A_{1\alpha_i}|\cdots|A_{m_i\alpha_i}$ there is at least $A_{t\alpha_i}$ with $k$-particles such that $A_{t\alpha_i}^{k_1}$ and $A_{t\alpha_i}^{k_2}$ are correlated ({\it labelled condition} (d')).  Here $b$ is a constant greater than 0, $k_1+k_2=k$, $A_{t{\alpha_i}}^{k_1}|A_{t{\alpha_i}}^{k_2}$ is any bipartition of $A_{t\alpha_i}$ containing $k$-particles, $A'=\{A_{1\alpha_i}|\cdots|A_{m_i\alpha_i}\}$ is the set
of all partitions satisfying the conditions (a), (b), (c'), and (d'), $s_k(A')$ stands for the cardinality of elements in the set $A'$.

Here we offer an example to elaborate the differences among these quantifications. Considering a 4-qubit $W$ state $|W_4\rangle=\frac{|1000\rangle+|0100\rangle+|0010\rangle+|0001\rangle}{2}$ and setting $p=2$, $k=3$, we can get  $E_{2-3}(|W_4\rangle)=\sqrt{\frac{3}{8}}$, $\mathcal{E}_{2-3}(|W_4\rangle)=\sqrt[28]{(\frac{3}{8})^5(\frac{5}{12})^6(\frac{1}{2})^3}$, $E'_{2-3}(|W_4\rangle)=\sqrt{\frac{3}{8}}$, and $\mathcal{E}'_{2-3}(|W_4\rangle)=\sqrt{\frac{3}{8}}$, respectively. Computing $E_{2-3}(|W_4\rangle)$ or $\mathcal{E}_{2-3}(|W_4\rangle)$, we need to consider fourteen partitions as per the previous definitions, whereas if we leverage the above strengthened restriction on partitions, only four partitions are required to be involved in Eqs. (\ref{14}) and (\ref{16}). This intuitively illustrates that the process of calculating the entanglement of a quantum state can be simplified by using the latter proposed quantifiers.

It is worth noting that the quantification forms in Definition 3 and Definition 4 obeying the properties: vanishing for any genuine $k$-producible pure state, being strictly positive for any non-genuine $k$-producible pure state. This implies that they can be used to distinguish whether a multipartite pure state is genuinely $k$-producible.

We show some specific $s'_k(n)$ in Table \ref{tab:t2}, where $n$ takes from 3 to 8 and $k$ goes from 1 to $n-1$. A comparison between Tables \ref{tab:t1} and \ref{tab:t2} shows that the quantifications defined in this section are computationally simpler when it comes to judge whether a quantum state is genuinely  $k$-producible.

{\bf Definition 5}. A mixed state $\rho$ with ensemble decomposition $\{p_i, |\phi_i\}$ is called genuinely strong $k$-producible if every $|\phi_i\rangle$ is genuinely $k$-producible.

Note that the concepts of genuine $k$-producibility and genuinely strong $k$-producibility are equivalent for pure states.

If Definition 3 and Definition 4 above are extended to mixed states using the convex roof extension, they will be effective quantifiers for determining whether a quantum state $\rho$ is genuinely strong $k$-producible. A quantum state is genuinely strong $k$-producible iff $E'_{p-k}(\rho)=\mathcal{E}'_{p-k}(\rho)=0$.

\section{Conclusion}\label{VII}
This paper provides a series of entanglement measures embarking on the two different main lines, one of which is to find the minimum of entanglement and the other is to take the geometric mean of entanglement. We show in detail that the $q$-$(k+1)$-(G)PE concurrence and $\alpha$-$(k+1)$-(G)PE concurrence serve as rational entanglement measures satisfying monotonicity, strong monotonicity, convexity, and vanishing iff the state is $k$-producible. In addition, we present the lower bounds on these measures by considering the PI part of quantum states and establish the relation between them. Furthermore, we compare these measures obtained by us and explain what they have in common and how they differ from one another. Besides, we define the quantification forms of non-genuinely strong $k$-producibility, which clearly distinguish genuinely strong $k$-producible states from other states and is convenient for calculation. These results may be available in practical applications and shed light on further investigations of multipartite quantum entanglement.

\section*{ACKNOWLEDGMENTS}
This work was supported by the National Natural Science Foundation of China under Grants No. 12071110 and No. 62271189, and the Hebei Central Guidance on Local Science and Technology Development Foundation of China under Grant No. 236Z7604G.

%\bibliographystyle{apsrev4-1}
%\bibliography{../Some_Tools/mybib}

\end{document}